\documentclass[aps,showpacs]{revtex4}
\usepackage{psfig}

\begin{document}
%
\title{
\[ \vspace{-2cm} \]
\noindent\hfill\hbox to 1.5in{\rm  } \vskip 1pt
\noindent\hfill\hbox to 1.5in{\rm SLAC-PUB-10478 \hfill  } \vskip
1pt \noindent\hfill\hbox to 1.5in{\rm June 2, 2004 \hfill}\vskip
10pt Cosmology Quantized in Cosmic Time\footnote{This work was
supported by the U.~S.~DOE, Contract No.~DE-AC03-76SF00515.}}
\author{Marvin Weinstein and Ratindranath Akhoury\footnote{On sabbatical leave from
Dept. of Physics, University of Michigan, Ann Arbor, MI 48109-1120, work supported
in part by the DOE}}
\address{Stanford Linear Accelerator Center, Stanford University,
  Stanford, California 94309}
\date{November 13, 2003}
\begin{abstract}
This paper discusses the problem of inflation in the
context of Friedmann-Robertson-Walker Cosmology.  We show how,
after a simple change of variables, to quantize
the problem in a way which parallels the classical discussion.
The result is that two of the Einstein equations
arise as exact equations of motion and one of the usual Einstein
equations (suitably quantized) survives as a constraint equation
to be imposed on the space of physical states.  However, the
Friedmann equation, which is also a constraint equation and which
is the basis of the Wheeler-deWitt
equation, acquires a welcome quantum correction that becomes significant
for small scale factors.  We discuss the extension of this
result to a full quantum mechanical derivation of the anisotropy
($\delta \rho /\rho$) in the cosmic microwave background radiation, and
the possibility that the extra term in the Friedmann equation could have
observable consequences.
To clarify the general formalism and explicitly show why we choose to weaken
the statement of the Wheeler-deWitt equation, we apply the general formalism
to de Sitter space.  After exactly solving the relevant Heisenberg equations
of motion we give a detailed discussion of the subtleties associated with defining
physical states and the emergence of the classical theory.  This computation
provides the striking result that quantum corrections to this long wavelength limit
of gravity eliminate the problem of the {\it big crunch\/}.  We also show that
the same corrections lead to possibly measurable effects on the CMB radiation.
For the sake of completeness, we discuss the special case, $\Lambda=0$, and
its relation to Minkowski space.
Finally, we suggest interesting ways in which these techniques
can be generalized to cast light on the question of chaotic
or eternal inflation.  In particular, we suggest one can put an
experimental lower bound on the distance to a universe with a
scale factor very different from our own, by looking
at its effects on our CMB radiation.
\end{abstract}
\maketitle

\newcommand{\ba}{\begin{eqnarray}}
\newcommand{\ea}{\end{eqnarray}}
\newcommand{\x}{\mbox{$\vec{x}$}}
\newcommand{\dphidt}{{\epsilon d\phi(t,\x) \over dt}}
\newcommand{\Phidot}{{d\Phi(t) \over dt}}
\newcommand{\Phiddot}{{d^2\Phi(t) \over dt^2}}
\newcommand{\adot}{{da(t) \over dt}}
\newcommand{\addot}{{d^2a(t) \over dt^2}}
\newcommand{\gradphi}{\epsilon\vec{\nabla}\phi(t,\x)}
\newcommand{\gradphisq}{\epsilon^2\vec{\nabla}\phi(t,\x)\cdot\vec{\nabla}\phi(t,\x)}
\newcommand{\be}{\begin{equation}}
\newcommand{\ee}{\end{equation}}
\newcommand{\goo}{(1 + 2\epsilon\,\chi(t,\vec{x}))}
\newcommand{\gxx}{(1 - 2\epsilon\,\chi(t,\vec{x}))}
\newcommand{\gooinv}{{1\over 1 + 2\epsilon\,\chi(t,\vec{x})}}
\newcommand{\gxxinv}{a(t)^2\,({1 - 2\epsilon\,\chi(t,\vec{x})})}
\newcommand{\udot}{{d u(t)\over dt}}
\newcommand{\uddot}{{d^2u(t) \over dt^2}}
\newcommand{\Hub}{{\cal H}}
\newcommand{\Hdot}{{d{\cal H}(t) \over dt}}
\newcommand{\Hddot}{{d^2{\cal H}(t) \over dt^2}}
\def\ket#1{\vert #1 \rangle}
\def\bra#1{\langle #1 \vert}
\def\vev#1{\left< #1 \right>}
\def\A{\frac{3\kappa^2}{16{\bf V}}}
\def\gop{ \frac{\kappa^2}{3}\left(\frac{1}{2}
    \left({d\Phi(t)\over dt}\right)^2 + V(\Phi) \right)}
\section{Introduction}

The COBE\cite{Mather:pc} and
WMAP\cite{Peiris:2003ff}\cite{Komatsu:2003fd}\cite{Nolta:2003uy}
measurements of the anisotropy in the cosmic microwave
background(CMB) radiation agree remarkably well the predictions of
slow-roll inflation\cite{Guth:prd23}.  This agreement provides a
strong reason to believe that the paradigm for computing the
fluctuations\cite{Mukhanov:PRpt215} in $\delta \rho /\rho$ is
correct. Perhaps the most striking feature of this result is that they
represent an imprinting of the
structure of the quantum state of the field theory, at the time
inflation begins, onto the electromagnetic radiation that comes to
us from the surface of last scattering. Unfortunately, derivations
of this effect usually mix classical and
quantum ideas and so, it is difficult to determine how
they would change given a fully quantum mechanical
treatment.  This paper fills this gap.  We begin by showing how, working in
fixed, co-moving coordinates, one can canonically quantize the
theory of the Friedmann-Robertson-Walker(FRW) metric,
\be
   ds^2 = - dt^2 + a(t)^2 d\vec{x}\cdot d\vec{x} ,
\label{FRW}
\ee
and the spatially constant part of the inflaton field, $\Phi(t)$,
in a straightforward manner.  We then show that the quantized system
has states for which the expectation values of the scale factor and
inflaton field satisfy the equations associated with the inflationary
scenario.  This, of course means that starting in one of these
states one can construct the usual perturbative analysis but, with the added
benefit, that the formalism will automatically generate terms responsible
for back reaction.

It is important to emphasize that our approach assumes that
getting quantum mechanics to describe the evolution of the system
in {\it cosmic time\/} is paramount. For this reason we find that
we cannot impose a strong form of the Wheeler-deWitt equation.  In
our formalism, geometry, which is defined by the condition that
the Einstein equations be true, is an emergent phenomenon.  It
exists only for some quantum states and then, only when the scale
factor becomes large.\cite{minisuperspace}  To clarify the subtle way in which this
works we begin by setting out the general formalism and then, we
exactly solve our equations for the case of de Sitter space. Next,
we identify a class of states which correspond to systems which,
at large times, behave both classically and in complete accord
with the full set of Einstein equations.  As we will show, a bonus
of this approach is that the quantum corrections to the Einstein
equations, which become important when the scale factor is small,
completely eliminate the problem of the {\t big crunch\/}.  In the
latter sections of this paper we discuss ways to extend this
derivation to compute possible experimental consequences of this
extra term.

We should note that some of the results presented in this paper were discussed
in two earlier preprints\cite{Weinstein:2003ya}.

\section{The Two Meanings of "Classical"}

There are, in fact, two ways in which current derivations of
${\delta \rho/\rho}$ invoke classical arguments.  First, these
derivations begin by treating both the scale factor $a(t)$ and the
spatially constant part of the inflaton field, $\Phi(t)$, as
classical time-dependent background fields. One then studies the
physics of the classical action \ba
    {\cal S} = {\bf V}\int dt \sqrt{-g} \left[ {R(t) \over 2\kappa^2}
    + {1\over 2}{d\Phi(t) \over dt}^2 - V(\Phi(t)) \right] .
\label{simpleaction}
\ea

The second appearance of classical ideas occurs when one adds back
spatially varying fluctuations in the Newtonian potential and the
inflaton field as quantum operators.  Many familiar derivations
then discuss the perturbative evolution of these fields in the
time-dependent background of the classical solution and, at an
appropriate point in the discussion, say ``and then the field goes
classical''. This much less important introduction of classical
ideas is used to convert the quantum computation of the two-point
correlation function for the density operator to an ensemble
average of gaussian fluctuations.  In reality, this statement is
just a way of avoiding any discussion of the physics of {\it
squeezed states\/} and {\it quantum non-demolition
variables\/}\cite{Polarski:1995jg}. While we do not discuss this
issue in this paper, we will return to it in a longer, more
pedagogical paper, which is in preparation.  This longer paper
will show how to extend the results presented here to a full
quantum treatment of $\delta \rho/\rho$. The point we wish to
emphasize at this juncture is that a full quantum treatment of the
spatially constant part of the problem, appropriately extended to
include the spatially varying modes of the fields to second order,
provides a complete quantum picture of all of the physics which
can be experimentally tested in the foreseeable future.

\section{The Classical Problem}

Before discussing our approach to the quantum treatment of FRW cosmology
it is important to demonstrate that the classical version of our formalism
does no violence to the usual Einstein theory.  We demonstrate this in
this section.  In the next section we show how to canonically quantize the same theory.

Simplifying the usual derivations of the Einstein equations for FRW cosmology
is easily accomplished if one observes that experimentally we are dealing
with a spatially flat universe and so it is perfectly adequate to formulate
the problem in a definite coordinate system.  In the discussion which follows,
we take this to be co-moving coordinates in which the metric takes the general
form shown in Eq.\ref{FRW}.

We already noted that, restricting attention to the classical problem
of a scalar field in an FRW cosmology, the action reduces to the form
shown in Eq.\ref{simpleaction}, where ${\bf V}$ is the volume of the
region in which the theory is being defined, $\sqrt{-g} = a(t)^3$ and
the scalar curvature times $\sqrt{-g}$ is given by
\be
    \sqrt{-g}\,R(t) = {3 \over \kappa^2}\,a(t) {da(t)\over dt}^2 + {3
    \over \kappa^2}\, a(t)^2 {d^2 a(t)\over dt^2} .
\ee
(Clearly, when we
generalize to the computation of $\delta\rho/\rho$, the volume, {\bf V},
must be taken to be larger than the horizon volume at the time of inflation
in order to avoid edge effects.)

Substituting these expressions into Eq.\ref{simpleaction} and integrating
by parts, to eliminate the term with $d^2 a(t)/dt^2$, we obtain
\be
{\cal S} = {\bf V} \int dt\,\left[ - {3 \over \kappa^2} a(t) \left({d a(t) \over dt}\right)^2
+ {1\over 2} a(t)^3 \left({ d\Phi(t) \over dt} \right)^2 - a(t)^3 V(\Phi(t)) \right] .
\label{FRWact}
\ee
Next, to simplify the analysis of the quantum version of this problem, we
make the change of variables $u(t)^2 = a(t)^{3}$, which leads to the action
\be
{\cal S} = {\bf V} \int dt\,\left[ - {4 \over 3\kappa^2} \left({d u(t) \over dt}\right)^2
+ {1\over 2} u(t)^2 \left({ d\Phi(t) \over dt} \right)^2 - u(t)^2 V(\Phi(t)) \right].
\ee
This change of variables merely simplifies the classical discussion, however it has a
greater significance for the quantized theory.  This is because we can choose
$-\infty \le u \le \infty$, whereas the only physically allowable range for $a$
is $ 0 \le a \le \infty$.  It is only for the space of square-integrable functions
on the interval $-\infty \le u \le \infty$ that the Heisenberg equations of motion
can be obtained by canonical manipulations.

There are only two Euler-Lagrange equations for this system:
\be
{8 \over 3\kappa^2} \uddot + 2u(t) \left( {1\over 2}\left(\Phidot\right)^2 - V(\Phi(t))\right)= 0
\label{EL1}
\quad {\rm and} \quad
-u(t)^2 \left(\Phiddot + 3{\cal H}(t) \Phidot + {d V(\Phi) \over d\Phi(t)} \right) = 0 ;
\label{EL2}
\ee
where the Hubble parameter, ${\cal H}$, is defined as
\be
{\cal H} = {1 \over a(t)} \adot = {2 \over 3 u(t)} \udot .
\ee
Thus, by quantizing in this fixed gauge, we fail to obtain
the full set of Einstein equations.  The missing equations are the Friedmann equation
and its time derivative
\be
    {\cal H}(t)^2 = {\kappa^2 \over 3} \left( {1\over 2} \left(\Phidot\right)^2
    + V(\Phi(t)) \right)
\quad {\rm and} \quad
    \Hdot = - {\kappa^2 \over 2} \left(\Phidot\right)^2 .
\label{Feq}
\ee

A sophisticated way of explaining why we fail to obtain these
equations is to note that by fixing the form of the metric to be
that given in Eq.\ref{FRW}, we have lost the freedom to vary the
lapse and shift functions.  But this is what we must do to obtain
the missing equations from a Lagrangian formulation.  This
predicament is not unique to gravity; it occurs in ordinary
electrodynamics if one chooses $A_0 = 0$ gauge.  As is well known,
in this gauge we obtain all of the Maxwell equations except
Coulomb's law, $\vec{\nabla}\cdot \vec{E} - \rho = 0$, as exact
equations of motion.  However, it follows from the equations we do
have, that Coulomb's law commutes with the evolution, {\it i.e.\/}
if we set it equal to zero it remains zero.  Hence, in this gauge,
while there are many solutions to the equations of motion, we can
select the ones we choose to call physical by imposing an extra
time-independent constraint.

The situation with the Friedmann equation and its time derivative
is analogous to the situation in electrodynamics. We will
now show that while Eqs.\ref{Feq}, are not equations of motion, if
they are imposed at any one time, then they will continue to be
true at all later times.  (In other words they are constraint
equations.)

To prove these constraints
are preserved by the equation of motion we begin by differentiating
${\cal H}$ with respect to $t$ to obtain
\be
\uddot = {3 u(t) \over 2}\left(\Hdot+ {3\over 2} {\cal H}(t)^2 \right).
\ee
Substituting this into Eq.\ref{EL1} and rearranging terms we obtain
\be
  {2 u(t)\over \kappa^2} \left( 2 \Hdot + 3 {\cal H}(t)^2 + \kappa^2 \left(\Phidot\right)^2
  -\kappa^2\left({1\over 2}\left(\Phidot\right)^2 + V(\Phi(t)) \right)\right) = 0 ,
\ee
which can be immediately rewritten in the form
\be
  {2 u(t) \over \kappa^2} \left[
  \left(2 \Hdot + \kappa^2 \left(\Phidot\right)^2\right)+ 3 \left({\cal H}(t)^2
  -{\kappa^2\over 3} \left({1\over 2}\left(\Phidot\right)^2
  + V(\Phi(t)) \right) \right)\right] = 0 .
\label{goodone}
\ee
If we, for convenience, define
\be
{\bf G} = {\cal H}(t)^2 - {\kappa^2 \over 3} \left( {1\over 2} \left(\Phidot\right)^2
    + V(\Phi(t)) \right) ,
\ee
the equation of motion for $\Phi(t)$ implies
\be
 {d {\bf G} \over dt} = 2 {\cal H}(t) \Hdot + \kappa^2 {\cal H}(t) \left(\Phidot\right)^2
 = 2{\cal H}(t) \left( \Hdot + {\kappa^2\over 2} \left(\Phidot\right)^2\right) .
\ee
The missing Einstein equations are equivalent to requiring
that both ${\bf G}$ and ${d{\bf G}/dt}$ vanish for all time.  Substituting these definitions
into Eq.\ref{goodone}, we obtain the exact equation of motion
\be
    {2 u(t) \over \kappa^2} \left( {1 \over {\cal H}(t)} \left(d{\bf G}\over dt\right)
        + 3 {\bf G} \right) =0 .
\label{constraintproof}
\ee
This equation shows that if, at time $t=t_0$, ${\bf G} = 0$, then the exact equation
of motion implies ${d {\bf G}/dt}$ will also vanish.  Given that this equation is
a first order differential equation for ${\bf G}(t)$, it follows that ${\bf G}(t)=0$
exactly.  In other words, we arrive at the desired result.  The Friedmann equation is, in
analogy to Coulomb's law in $A_0=0$ gauge, a constraint which can be imposed at
a single time and which will continue to be true at all later times.  We will show
that a similar theorem can be proven for the quantum theory; however, we will argue that
the analogy with QED is not perfect.

Before moving on to the quantum theory, let us spend a few moments
discussing the Hamiltonian version of the classical theory. We do this
to show why it is possible to confuse the Friedmann equation
with the Hamiltonian at the classical level .

Following the usual
prescription, we vary Eq.\ref{FRWact} with respect to $du/dt$ and
$d\Phi/dt$ to obtain
\be
  p_u = -{\bf V} {8  \over 3 \kappa^2} \udot
  \quad;\quad p_\Phi = {\bf V} u^2 \Phidot .
\ee
We then construct the Hamiltonian
\be
{\bf H} = p_u \udot + p_\Phi \Phidot - {\cal L}
= -{3\kappa^2 \over 16 {\bf V}} p_u^2 + {1 \over  2{\bf V} u^2 }p_\phi^2
+ {\bf V} u^2 V(\Phi)
\ee

An important feature of this Hamiltonian is that
due to the minus sign in front of the $p_\mu^2$ term,
it has no minimum.  Fortunately, this doesn't matter.
To see this, we simply rewrite the Hamiltonian in terms of
$du/dt$ and $d\Phi/dt$.  This leads to the expression
\be
{\bf H} = {\bf V}\left[ -{4\over 3\kappa^2} \left(\udot\right)^2 + u^2 \left( {1\over 2}
\left(\Phidot\right)^2 + V(\Phi) \right) \right] .
\ee
Substituting the definition of ${\cal H}$, this becomes
\ba
{\bf H} &=& -{\bf V} u^2 \left[ {4\over 3\kappa^2} {1\over u^2} \left(\udot \right)^2
-\left( {1\over 2} \left(\Phidot\right)^2+V(\Phi)\right)\right]
= -{\bf V} u^2\left[ {3 {\cal H}^2 \over \kappa^2 }
- \left( {1\over 2} \left(\Phidot\right)^2+V(\Phi)\right)\right] \nonumber\\
{\bf H} &=& -{\bf V} {3 u^2\over \kappa^2} {\bf G} .
\ea
This shows that the Hamiltonian, {\bf H}, is proportional to the constraint, ${\bf G}$.
It follows that setting ${\bf G}=0$ means ${\bf H}=0$, which tells us that
the Hamiltonian vanishes for physical solutions.  In other words, if we start a system out
at $t=t_0$ in a configuration which has zero energy, it will stay at zero energy
and never explore the region of arbitrarily negative energy.
The identification of the Hamiltonian with the constraint equation
is the content of the Wheeler-DeWitt equation.

\section{Canonical Quantization of the Theory}

Now that we have seen that our formalism, including the change of variables from
$a(t)$ to $u(t)$, does no violence to the classical theory, we will proceed to a discussion
of the quantum mechanics.

Starting from the classical Lagrangian, we define the quantum Hamiltonian
\be
{\bf H} = - {3 \kappa^2 \over 16 {\bf V} } p_u^2 + {1\over 2 {\bf V} u^2} p_\Phi^2 + {\bf V} u^2 V(\Phi)
\ee
where the operators $u$,$\Phi$ and their conjugate momenta have the commutation relations
\be
\left[p_u, u\right] = -i \quad;\quad \left[p_\Phi,\Phi\right]=-i .
\ee
All other commutators vanish.
To derive the Heisenberg equations of motion, note that for any operator ${\bf O}$,
the Heisenberg operator is $O(t) = e^{i{\bf H}t} {\bf O} e^{-i{\bf H}t} $.
Commuting ${\bf H}$ with the operators $u$ and $\Phi$, we obtain
\ba
 \udot &=& i\, \left[{\bf H},u\right] = -{3\kappa^2\over 8 {\bf V}} p_u \nonumber\\
 \Phidot &=& i\,\left[{\bf H},\Phi\right] = {1\over u^2 {\bf V}} p_\Phi \nonumber\\
 \uddot &=& i\,\left[{\bf H},\udot\right] =
 -{3\kappa^2\over 4} u \left[ {1\over 2} \left(\Phidot\right)^2 - V(\Phi)\right]\nonumber\\
 \Phiddot &=& {3\kappa^2 \over 16 {\bf V}} \left( {1\over u^2} p_u {1\over u}
 + {1\over u} p_u {1\over u^2} + p_u {1\over u^3} + {1\over u^3}p_u \right)p_\Phi
 - {dV(\Phi) \over d\Phi} .
\label{heisenone}
\ea
This shows that the two dynamical equations of the classical
theory are also exact operator equations of motion in the quantum theory.
What is missing, as in the classical theory, are the constraint equations.
In order to find the constraint equations that commute with
the Hamiltonian, we begin by rewriting the equation for $\Phi$ in the
suggestive form
\be
    \Phiddot + 3 {\cal H} \Phidot + {dV(\Phi) \over d\Phi} = 0 ,
\label{heisentwo}
\ee
where the quantum version of the  {\it Hubble} operator ${\cal H}$ is perforce
\be
    {\cal H} = -{\kappa^2 \over 8{\bf V}}\left( p_u {1\over u} + {1\over u^3} p_u u^2 \right) .
\label{Hubbleop}
\ee
Next we compute its time derivative from the equation
\be
    {d{\cal H}\over dt} = i\,\left[{\bf H},{\cal H}\right] .
\ee
Finally, to find the quantum version of the conserved constraint operator, ${\bf G}$,
we follow the  classical procedure and write
\be
\uddot = {3 u\over 2}\left( {d{\cal H}\over dt} + {3\over 2} {\cal H}^2
  -{9 \kappa^4 \over 128 {\bf V}^2 u^4} \right) .
\label{extra}
\ee
The extra term is the quantum correction to the classical formula.  It is obtained by
explicitly taking the difference between the expression for $d^2u/dt^2$ and the combination
$(3 u/ 2)\left( {d{\cal H}/dt} + 3{\cal H}^2/2 \right)$.
(This step involves commutator gymnastics better left to Maple.)
Once again, paralleling the classical discussion, we substitute the expression
for $d^2u/dt^2$ into the Heisenberg equation of motion for $u$,
obtaining
\be
{3 u \over 2} \left( {d{\cal H}\over dt} + {3\over 2}{\cal H}^2
 -{9 \kappa^4 \over 128 {\bf V}^2 u^4}\right)
 + {3\kappa^2 u\over 4}\left(\left(\Phidot\right)^2
 - \left({1\over 2}\left(\Phidot\right)^2 + V(\Phi)\right)\right) = 0.
 \label{eqa}
 \ee
At this point it is tempting to parallel the classical discussion and define the
operator
\be
{\bf G} = \Hub^2 -{\kappa^2\over 3}\left(\frac{1}{2}\left(\Phidot\right)^2 + V(\Phi)\right)
+ {\bf Q} ,
\ee
where
\be
{\bf Q}=-{3\kappa^4\over 64 {\bf V}^2 u^4} ,
\label{Qop}
\ee
and show that if it annihilates a state at any one time, then it annihilates it for
all times.  We will now show that this can be done, however we will then argue that
identifying the kernel of this operator with the space of physical states is
incorrect.  To proceed with the proof substitute this definion into
Eq.\ref{eqa} to obtain the operator equation of motion
\be
    {3u\over 4} \left(2 \Hdot + \kappa^2 \left(\Phidot\right)^2 + 3 {\bf G}\right) =0 .
\label{almost}
\ee
This is almost what we need to show that the space of physical states, defined to be those
which obey the condition ${\bf G}(t)\ket{\psi}=0$, is invariant under Hamiltonian evolution.
Clearly, we will be able to use Eq.\ref{almost} to complete the proof
if we can show that there exists an operator ${\bf A}$ such that
\be
{d{\bf G}\over dt} = {\bf A} \left( 2 \Hdot + \kappa^2 \left( \Phidot \right)^2\right) .
\label{gdoteq}
\ee
To find ${\bf A}$, explicitly compute
\ba
{d{\bf G}\over dt} &=& i\,\left[{\bf H},{\bf G}\right] \nonumber\\
&=& \Hub \Hdot + \Hdot \Hub + \kappa^2 \Hub \left(\Phidot\right)^2 + \left[ {\bf H},{\bf Q}\right]\nonumber\\
&=& \Hub \left(2 \Hdot + \kappa^2 \left(\Phidot\right)^2 \right)+ \left[\Hdot,\Hub\right] + \left[ {\bf H},{\bf Q}\right] ,
\ea
and substitute this result into Eq.\ref{gdoteq}.  The resulting equation can then be
rearranged into the form
\be
\left({\bf A} - \Hub\right) \left( 2\Hdot + \kappa^2\left(\Phidot\right)^2\right) =
\left[ {\bf H},{\bf Q}\right] + \left[\Hdot,\Hub\right] .
\ee
Solving this equation for ${\bf A}$, we obtain
\be
    {\bf A} = \Hub + \left(\left[ {\bf H},{\bf Q}\right] + \left[\Hdot,\Hub\right]\right)
    \left( 2\Hdot + \kappa^2 \left(\Phidot\right)^2\right)^{-1}
 ,
\ee
which allows us to rewrite the Heisenberg equation of motion for $u$ as
\be
{3 u\over 4} \left({1\over {\bf A}} {d {\bf G}\over dt} + 3 {\bf G}\right) =0 .
\label{gaugeinv}
\ee
Given that Eq.\ref{gaugeinv} is an exact operator equation of motion, we see that
if we could define the space of states by the condition ${\bf G(t_0)}\ket{\psi}=0$,
then Eq.\ref{gaugeinv} proves that this condition will hold for all time.
Note, however, that given this definition of ${\bf G}$, it follows immediately that
\be
    {\bf G}(t) = -\frac{\kappa^2}{3{\bf V} u(t)^2}\,{\bf H}  .
\ee Thus, while we can define the space of physical states, to be
those which are annihilated by the Hamiltonian, obviously this
immediately leads to a contradiction between the Schroedinger and
Heisenberg picture.  This is because $H\ket{\psi}=0$ implies that
the state does not evolve in the Schroedinger picture, whereas we
have already shown that the operators $u(t)$ and $\Phi(t)$ do
evolve in time.

\section{A Better State Condition}

One way to avoid the inconsistency between the Schroedinger picture and Heisenberg
equations of motion is to observe that we can define a one-parameter family
of possible gauge-conditions as follows:
\be
    {\bf G}_\alpha = {\cal H}^2 - \gop + \alpha {\bf Q}
\ee
Then, following the previous arguments, we can show that any of these ${\bf G}_\alpha$ also
satisfies an  equation of the form
\be
{3 u\over 4} \left({1\over {\bf A_\alpha}} {d {\bf G_\alpha}\over dt} + 3 {\bf G_\alpha}\right) =0.
\label{gaugeinvtwo}
\ee
This means that a state is annihilated by ${\bf G}_\alpha(t)$ at any one time, $t_0$, will be
annihilated by ${\bf G}_\alpha(t)$ for all times.  Since, independent of the value of
$\alpha$,  the extra pieces in all of these modified Friedmann
equations vanish for large $u(t)$ it follows that, as before,
in the limit of large $u(t)$ the expectation values of the dynamical fields will satisfy
all of the Einstein equations.

To see that these alternative forms of the gauge-condition
avoid direct conflict between the Schroedinger and Heisenberg picture simply
substitute the explicit form of the Hubble operator, Eq.\ref{Hubbleop},
and the definition of ${\bf Q}$, Eq.\ref{Qop}, into the definition of ${\bf G}_{\alpha}$
and rewrite it as
\ba
    {\bf G}_\alpha &=& {\cal H}^2+ \alpha {\bf Q} - \gop \nonumber\\
    &=& \frac{\kappa^4}{16 {\bf V}^2 u^2} p_u^2 + (1-\alpha){3\kappa^4\over 64 {\bf V}^2 u^4}
    - \gop  \nonumber\\
    &=& -\frac{\kappa^2}{3 {\bf V} u^2}{\bf H} + (1-\alpha){3\kappa^4\over 64 {\bf V}^2 u^4} .
\label{galpha}
\ea
Now, since the Hamiltonian, ${\bf H}$, is time independent, we see that
\be
    {\bf G}_\alpha(t) = -\frac{\kappa^2}{3{\bf V}u(t)^2} \,\left[{\bf H}
    - (1-\alpha){9\kappa^2\over 64 {\bf V} u(t)^2} \right] .
\ee
Thus, it is only for $\alpha=1$ that ${\bf G}_\alpha\ket{\Psi}=0$
implies that the Hamiltonian annihilates the state.

Unfortunately, as we will see in the next section, in exactly solvable models we can
explicitly show that the solutions to the equation ${\bf G}_\alpha\ket{\psi}=0$ are
not normalizable, and attempting to impose this strong condition for any $\alpha$ leads to
problems interpreting the quantum mechanical theory.  For this reason we propose a weak form
of the condition, namely: a state is physical if
\be
    \lim_{t\rightarrow \pm\infty} {\bf G}(t) \ket{\Psi} = 0.
\label{weakasymp}
\ee
It should be clear from the fact that ${\bf Q}$ vanishes for large $t$ that Eq.\ref{weakasymp}
guarantees that geometry, in the sense that the familiar Einstein equations become
arbitrarily accurate, emerges dynamically at large times.  Another way of characterizing the
difference between this approach and the Wheeler-DeWitt equation is that, for
reasons which will be apparent in the next section, the underlying physics is more
closely related to a scattering problem, rather than an eigenvalue problem.

In the next sections, where we discuss the exact solution of de Sitter space,
we show that this asymptotic condition is satisfied for a large class of states.
Furthermore, the exact solution demonstrates why imposing
a stronger condition on physical states is neither necessary nor desirable.

\section{de Sitter Space: An Exactly Solvable Problem}

Since our assertion that it is unnecessary to adopt a strong version of the gauge condition
flies in the face of conventional wisdom,
it is important to show how things work in an exactly solvable example.
For this reason we devote the next few sections of this paper to a discussion
of de Sitter space.

Begin by considering the general action of the FRW problem, but with
$V(\Phi)$ replaced by a cosmological constant $\Lambda$, so that the Hamiltonian
takes the form
\be
{\bf H} = - {3 \kappa^2 \over 16 {\bf V} } p_u^2 + {1\over 2 {\bf V} u^2} p_\Phi^2
+ {\bf V} u^2 \Lambda
\label{vphizero}
\ee
We then note that, since the conjugate variable to $p_\Phi$ doesn't appear in the Hamiltonian,
we are free to work in sectors of the Hilbert space in which $p_\Phi$ takes a definite
value.  For the particular sector defined by the condition $p_\Phi \ket{\psi} = 0$
the Hamiltonian takes the simpler form
\be
{\bf H} = - {3 \kappa^2 \over 16 {\bf V} } p_u^2 +  {\bf V} u^2 \Lambda,
\label{dsham}
\ee
which we immediately recognize as a theory with a cosmological constant, whose
solution at the classical level is just de Sitter space.

Direct commutation of the Hamiltonian, Eq.\ref{dsham},  with the operators
$u(t)$ and $p_u(t)$ yields the following Heisenberg equations of motion for
$u(t)$ and $p_u(t)$:
\be
    {du(t)\over dt} = -\frac{3\kappa^2}{8{\bf V}} p_u \,;\,
    {d^2u(t) \over dt^2} = \frac{3\kappa^2 \Lambda}{4} u .
\ee
The exact solution to these equations, written in terms of the operators $u(t=0)=u$ and $p_u(t=0)=p_u$
are
\ba
u(t) &=& \cosh(\omega t) u - \frac{3\kappa^2}{8 {\bf V} \omega} \sinh(\omega t) p_u
\nonumber\\
p_u(t) &=& \cosh(\omega t) p_u -\frac{8 {\bf V} \omega}{3\kappa^2}\sinh(\omega t) u ,
\label{exsols}
\ea
where we have defined
\be
    \omega = \sqrt{3\kappa^2 \Lambda \over 4} .
\ee

It is convenient to rewrite Eq.\ref{exsols} in terms of exponentials; i.e.,
\be
u(t) = \frac{e^{\omega t}}{2}\left( u - \frac{3\kappa^2}{8 {\bf V} \omega} p_u \right)+
\frac{3\kappa^2 e^{-\omega t}}{16 {\bf V} \omega}\,
\left( p_u + \frac{8 {\bf V}\omega}{3\kappa^2} u \right)
\ee
and to introduce the canonically conjugate asymptotic operators
\be
 u_\infty = \frac{1}{\sqrt{2}} \left( u-\frac{3\kappa^2}{8{\bf V}\omega}p_u
   \right) \,;\,
p_\infty = \frac{1}{\sqrt{2}}\left( p_u + \frac{8{\bf V} \omega}{3\kappa^2} u \right) .
\ee
In terms of these operators the solution for the operator $u(t)$ and the Hamiltonian take
the simple forms
\be
u(t) = \frac{1}{\sqrt{2}}\,e^{\omega t} u_\infty + \frac{1}{\sqrt{2}}\,
\frac{3\kappa^2}{8 {\bf V} \omega}\, e^{-\omega t} p_\infty ,
\label{exactu}
\ee
and
\be
{\bf H} = \frac{\sqrt{3\Lambda}\kappa}{4}\left( u_\infty p_\infty
+  p_\infty u_\infty \right) .
\label{hamiltonian}
\ee
Eq.\ref{exactu} shows why, in the preceding section, we stated that the underlying
physics is more closely related to the physics of a scattering problem than an
eigenvalue problem.  To establish the parallel all we need do is identify the
{\it in}-states with the eigenstates of $p_\infty$ and the {\it out}-states
with the eigenstates of $u_\infty$.  Note, since the Hamiltonian is time-independent
(Eq.\ref{hamiltonian} ), the expectation value of the energy in any state is perforce
time independent too.

From this point on all of the technical work is finished, the only chore which remains
is to extract the physical significance of these results.

\section{More About Physical States}

Before discussing the physical states of the quantum theory,
it is worth spending a few moments considering what the preceding results
mean in the context of the classical theory.  Obviously, Eqs.\ref{exactu} and
\ref{hamiltonian} are equally true for both the classical and
quantum versions of the theory; the only
difference between these cases being is that in the classical theory $u_\infty$ and
$p_\infty$ are simply numbers, whereas in the quantum theory they
are non-commuting operators.  Thus, for the classical
theory, imposing the condition that the energy vanishes is the same
as requiring either $u_\infty$ or $p_\infty$ to vanish.  This is,
of course, just the usual result: i.e., for the case of a
cosmological constant, the full, non-linear, set of Einstein
equations, admit only an expanding, or contracting, solution for
$a(t)$ or $u(t)$.  This is why running the expanding solution back in time
(or the contracting solution forward in time) always leads to a
{\it big crunch\/}.

Clearly, the situation is different for the quantum theory since
it is not possible to simply set an operator to zero.  If one
chooses the gauge-condition which corresponds to $\alpha=1$, i.e.
the Wheeler-deWitt equation, then one is looking for states
annihilated by the Hamiltonian.  Given that we can write
$p_\infty=-i\,{d \over d u_\infty}$, for a function of the form
$\ket{\psi}=e^{S(u_\infty)}$,this equation takes the simple form
\be
    {2\, u_\infty} {d S(u_\infty) \over du_\infty} = -1 ,
\ee
which has the solution
\be
    S(u_\infty) = - \ln(\sqrt{u_\infty}).
\ee
This of course means that $\ket{\psi}$ is of the form
\be
    \ket{\psi} \approx {1 \over \sqrt{u_\infty}}
\ee
which is not normalizable.  The situation is no better if one
chooses one of the gauge-conditions for which $\alpha \ne 1$.
It is because working with these non-normalizable states makes
interpreting the quantum theory so problematic that we adopt the
weaker asymptotic condition defined in Eq.\ref{weakasymp}

Intuitively, given the exact solution for $u(t)$, we see that any
state for which ${\bf H}\ket{\Psi}$ has a finite norm will, for
sufficiently large $|t|$, satisfy Eq.\ref{weakasymp} to arbitrary
accuracy. This means that essentially any Gaussian wave packet in $u_\infty$
will be a physical state. It also means that for large times all
the physics measured in such a state will be compatible with the
full set of Einstein equations. (Actually there is the additional
requirement that the wave-function, ${\bf H}\ket{\psi}$, vanishes
sufficiently rapidly at zero when written as a function $u_\infty$
or $p_\infty$. This, however, can be accomplished by multiplying
any shifted Gaussian in $u_\infty$ by an appropriate polynomial
in $u_\infty$.  This subtlety will not seriously affect the considerations
of the sections to follow and so we will ignore it.  It is, however,
important when we consider more complicated situations.)

\section{Quantum Histories}

Now that we have settled upon shifted Gaussian wavepackets as
good candidates for physical states, we turn
to a discussion of the only two physical observables in this
theory; the expansion rate and the volume of the universe.
Since we are working in the Heisenberg picture, where the choice of
state determines the entire subsequent evolution of the system,
we will henceforth refer to the choice of an allowed quantum state
as a choice of {\it quantum history\/}.
What we wish to ascertain is to what degree the value of each of the
observables depends upon the specific choice of {\it quantum history\/}.

Obviously, the exact solution given in Eq.\ref{exactu} shows that,
at large times, the expansion rate is attached to the scale factor
and is totally independent of the state.  This, however, is not
true of the volume. Thus, in the remainder of this section we will
discuss the degree to which the measured properties of the volume
operator differ from quantum history to quantum history.

Since we started off quantizing in a volume with coordinate size
${\bf V}$, the volume of the universe at any time is given by
\ba
    V(t) &=& {\bf V} u(t)^2 \cr
    &=& \frac{{\bf V}}{2}\, \left[
    e^{2\omega t} u_\infty^2
    + \left( \frac{3\kappa^2}{8{\bf V} \omega} \right)^2 \,e^{-2\omega t}
    p_\infty^2
    + \frac{3\kappa^2}{8{\bf V} \omega} \left(u_\infty p_\infty
    + p_\infty u_\infty\right)\right] . \label{voft}
\ea
A surprising feature of this formula is that for large
times in the past and future the volume operator $V(t)$ behaves
classically.  By this we mean that, if one measures $V(t)$ at some
late time, $t_1$, and obtain a definite value, then we will be
able to predict the value we will obtain if we measure $V(t)$ at
some later time $t_2$.  To see that this is the case we note that
Eq.\ref{voft} tells us that, for very large positive times, $V(t)$ is,
to arbitrarily high accuracy, proportional to the single operator
$u_\infty^2$ (at large negative times it is proportional to
$p_\infty^2$).  Thus we see that a measurement of $V(t_1)$, for
sufficiently large $t_1$, corresponds to a measurement of $u_\infty^2$,
which means that we know $V(t)$ for all times $t_2 > t_1$.

From the fact that $u_\infty$ and $p_\infty$ are canonically
conjugate variables we see that if we were to try and identify a
quantum history with an eigenstate of $p_\infty$, then the volume
operator would be well-determined in the past, but completely
undetermined in the future.  Conversely, eigenstates of $u_\infty$
correspond to states for which the volume operator is completely
well determined in the future, but completely undetermined in the
past.  Fortunately, the condition that physical states must be
normalizable states for which $\bra{\psi}{\bf H}^2\ket{\psi} <
\infty$ is true, tells us that we cannot identify such states with
quantum histories. States which can be identified with admissible
quantum histories are Gaussian packets of the form, \be \ket{\Psi}
= e^{-\frac{\gamma}{2} u_\infty^2} \ee and the coherent states,
$\ket{u_0,p_0,\gamma}$, obtained from them. These coherent states
are defined by \be
    \ket{u_0,p_0,\gamma} = e^{i p_0 u_\infty}\,e^{-i u_0 p_\infty} \ket{\Psi},
\ee
and the expectation values of $u_\infty$ and $p_\infty$ in these states are given by
\be
\bra{u_0,p_0,\gamma} u_\infty \ket{u_0,p_0,\gamma} = u_0 ,\,
\bra{u_0,p_0,\gamma} p_\infty \ket{u_0,p_0,\gamma} = p_0.
\ee
Moreover, the relevant products of these operators have the values
\ba
\bra{u_0,p_0,\gamma} u_\infty^2 \ket{u_0,p_0,\gamma} &=& u_0^2 + \frac{1}{2\gamma},
\nonumber\cr
\bra{u_0,p_0,\gamma} p_\infty^2 \ket{u_0,p_0,\gamma} &=& p_0^2 + \frac{\gamma}{2}, \nonumber\cr
\bra{u_0,p_0,\gamma} u_\infty p_\infty + p_\infty u_\infty\ket{u_0,p_0,\gamma}
&=& 2\Re(\vev{u_\infty\,p_\infty}) = 2\,u_0 p_0  .
\ea
The nice thing about such coherent states is that
they are the kind of states we would expect to obtain if, in the past,
we make a measurement which determines $V(-t)$ to have a central value
$\frac{{\bf V}}{2}\,e^{\omega|t|} p_0^2$, with a width parameterized by $\gamma$.
For this same packet, measurements of $V(t)$ in the distant future will produce
results centered about the value $\frac{{\bf V}}{2}\,e^{\omega|t|} u_0^2$,
with a width parameterized by $1/\gamma$.

\section{Equivalence Classes of Histories}

From this point on we will restrict the term quantum history to mean a coherent state
of the form defined above.  What we wish to show next is that
many of these histories are equivalent to one another in a way which we will make
precise.  Begin by considering
\be
    \vev{V(t)} = \bra{u_0,p_0,\gamma} V(t) \ket{u_0,p_0,\gamma}
    = \frac{{\bf V}}{2} \left[ e^{2\omega t} \vev{u_\infty^2}
    + \left( \frac{3\kappa^2}{8{\bf V} \omega} \right)^2 \,
    e^{-2\omega t} \vev{p_\infty^2} + \frac{3\kappa^2}{8{\bf V}\omega}\left(
    2\Re(\vev{u_\infty p_\infty})\right) \right] .
\label{vevoft}
\ee

It is obvious from Eq.\ref{vevoft} that at large times the volume
behaves as a single exponential, as expected from the solution of
the classical Einstein equations. More interesting, however, is
the fact that letting $t \rightarrow t+t_0$, where $t_0$ is
defined by the condition
\be
    e^{2\omega t_0} = {3\kappa^2 \over 8 {\bf V} \omega}
    \sqrt{\vev{p_\infty^2} \over \vev{u_\infty^2}} ,
\ee
allows us to rewrite Eq.\ref{vevoft} as
\ba
    \vev{V(t)}&=&\frac{3\kappa^2 \sqrt{\vev{u_\infty^2} \vev{p_\infty^2}}}{
    8 \omega} \left[ \cosh(\omega t)
    + \frac{\Re(\vev{u_\infty p_\infty})}{\sqrt{\vev{u_\infty^2} \vev{p_\infty^2}}}
    \right] \nonumber\\
    &=& \frac{\kappa^2\sqrt{\vev{u_\infty^2} \vev{p_\infty^2}}}{4 {\cal H}}
    \left[ \cosh(\omega t)
    + \frac{\Re(\vev{u_\infty p_\infty})}{\sqrt{\vev{u_\infty^2} \vev{p_\infty^2}}}
    \right]
\label{vevtwo}
\ea
Thus, we see $\vev{V(t)}$ corresponds to a system which
is contracting at large times in the past and which then bounces and begins
to re-expand in the future.  During most of this history the system
satisfies the Friedmann equation to high accuracy and expands
(or contracts) with a Hubble constant equal to
\be
    {\cal H} = \frac{2}{3} \omega = \sqrt{\frac{\kappa^2
\Lambda}{3}} .
\ee
However, there is a period in time where the
quantum corrections to the Friedmann equation dominate the
behavior; namely, at times $t \approx 1/\omega$.  Assuming, for
the sake of argument, that were to set $1/\kappa {\cal H} \approx
10^3$, as it is in many models of slow roll inflation, and
assuming $\sqrt{\vev{u_\infty^2} \vev{p_\infty^2}}$ to be of order
unity, then the minimum volume of the universe at the time of the
bounce is on the order of $10^3$ Planck volumes; i.e., on the
order to $10$ Planck-lengths in each dimension.  This sets the
order of magnitude of the scale at which the quantum corrections
become important.   It is gratifying that these quantum
corrections keep the system from contracting forever and ending in
a {\it big crunch\/}.

Another very interesting feature of Eq.\ref{vevtwo} is that it is
characterized by only two numbers, $\sqrt{\vev{u_\infty^2} \vev{p_\infty^2}}$ and
$\Re{\vev{u_\infty p_\infty}}/\sqrt{\vev{u_\infty^2} \vev{p_\infty^2}}$.
The first number is unrestricted in magnitude and roughly determines
the physical volume of the universe at the time of the bounce.
The second number, is constrained by the Schwarz inequality
to lie between plus and minus one, and parameterizes the degree to which
the behavior of the system during the time of the bounce deviates from
a pure hyperbolic cosine.  If the time over which the deviation takes place is
characterized by $1/\omega \approx 1/{\cal H}$, then the minimum size to which
the system contracts is characterized by the ratio of the energy density in the
state to the cosmological constant.  This statement follows from taking the
expectation value of the Hamiltonian as written in Eq.\ref{hamiltonian}, which
implies
\be
    \Re(\vev{u_\infty p_\infty}) = \frac{2}{\kappa \sqrt{3\Lambda}} \vev{{\bf H}} .
\ee
Note, it would appear from the Schwarz inequality that in principle one
could have a history for which the universe actually shrinks to zero
size before it bounces.  Fortunately it is easy to see that this can only
occur if $u_0$ or $p_0$ diverges, which violates the condition on allowable
physical states, since such states would have infinite values for
$\vev{{\bf H}^2}$.

Finally, Eq.\ref{vevtwo} shows that any two quantum histories which give the
same values for  $\sqrt{\vev{u_\infty^2} \vev{p_\infty^2}}$ and
$\Re(\vev{u_\infty p_\infty})\sqrt{\vev{u_\infty^2} \vev{p_\infty^2}}$,
see the same physics.  They only differ by the time at which they see the bounce
occur.  For Gaussian packets we see that this will be true for states which
are related by the transformation
\be
u_0 \rightarrow \lambda u_0, \,
p_0 \rightarrow \frac{p_0}{\lambda}, \, {\rm and \ }
\gamma \rightarrow \frac{\gamma}{\lambda^2} .
\ee
It is easy to check that this can be implemented by
a unitary transformation.  The values of $u_0$ and $p_0$ can be changed
by means of the shift operators used to define the coherent states in the
first place.  The width of the Gaussian can be changed by application of
a unitary {\it squeezing operator\/} of the form
\be
    e^{\left(\alpha(\gamma)\, {a^\dag}^2 - \alpha(\gamma)^\ast\, a^2\right)} ,
\ee
where the creation and annihilation operators are defined such that
\be
    u_\infty = \frac{1}{\sqrt{2\gamma}}\left(a^\dag + a\right) \quad {\rm and } \quad p_\infty = -i \,
    \sqrt{\frac{\gamma}{2}} \left(a^\dag - a\right) .
\ee

\section{Minkowski Space  $\Lambda=0$}

Finally, we would like to discuss what happens when we
take $\Lambda = 0$, because, in this case, things work quite a bit differently.
The $\Lambda=0$ Hamiltonian is
\be
    {\bf H} = -\,\frac{3\kappa^2}{16{\bf V}} p_u^2
\ee
and the Heisenberg equations of motion take the form
\be
    {d u \over dt}(t) = -\,\frac{3\kappa^2}{16{\bf V}} p_u  \quad ; \quad
    {d p_u \over dt}(t) = 0 .
\ee
The exact solution to these equations is
\be
    u(t)= u - \A\,p_u\,t   \quad ; \quad  p_u(t) = p_u
\label{heistwo}
\ee
Taking the square of $u(t)$ we obtain the volume operator
\be
    V(t) = {\bf V} u^2(t) = {\bf V}\left[ u^2 - \A\,\left(u\,p_u
    + p_u\,u\right)\,t + \left(\A\right)^2\,p_u^2\,t^2
\right] .
\ee
It follows once again that, as in the de Sitter case, the volume
operator becomes classical at large times in the past and the future.
In this case however there is a state which, while non-normalizable, satisfies the
condition $G(t) \ket{\Psi} = 0$ for all times; namely, the eigenstate of
$p_u$ with eigenvalue $0$.  Now, however, this condition is consistent with
the Heisenberg equations of motion, because in this eigenstate $u(t)=u$ and is
independent of time.  Moreover, this state satisfies
the requirement that $\vev{{\bf H}^2}$ is finite.
Obviously, this state is the limit of sequence
Gaussian packets in $p_u$ of smaller and smaller width.
If we choose this quantum history then, after we absorb the scale factor into
$\vec{x}$, we find that this history corresponds to a time-independent Minkowski space.

It is interesting to ask what other, less special, histories correspond to.
Let us assume we are working with an arbitrary coherent state of the form
discussed in the previous section.  Then, the expectation value
of the volume operator is
\be
    \vev{V(t)} = {\bf V} \left[\vev{u^2} - 2\A\,\Re(\vev{u\,p_u})\,t
    + \left(\A\right)^2\vev{p_u^2}\,t^2 \right] ,
\ee
which can be rewritten in the form
\be
\vev{V(t)} = {\bf V} \left[\frac{\vev{u^2}\vev{p_u^2}-\Re(\vev{u\,p_u})^2}{\vev{p_u^2}}
+\left(\frac{3\kappa^2}{16 {\bf V}}\right)^2 \,\vev{p_u^2}
\left(t - \frac{16{\bf V}\Re(\vev{u\,p_u})}{3\kappa^2\,\vev{p_u^2}}
\right)^2\right] .
\ee
Thus, we see that for the generic history, the case of zero cosmological
constant actually corresponds to a universe for which the volume factor
is expanding like $t^2$, or for which the scale factor $a(t)$ is growing
like $t^{2/3}$.  Surprisingly this corresponds to a universe dominated by
non-relativistic matter.  In other words, a non-vanishing energy
density present in the quantum excitations of the scale factor
produce the same effect as cold matter.

A final point worth mentioning is that, as in the case of de Sitter space,
the Schwarz inequality guarantees that the volume never shrinks to
zero for any allowable physical state; i.e., we never are in the situation of
a {\it big crunch}.  It is interesting to note that in this formalism the big
crunch is averted due to the quantum physics of the long wavelength modes of
the gravitational field and not short distance physics.

\section{Recovering the Classical Theory}

To this point we have set up our general formalism and we have shown how
things work for the exactly solvable case of de Sitter space, defined by
the two conditions $V(\Phi)=\Lambda$ and $p_\phi=0$.  The question which comes up
at this juncture is how things work in the more general setting when
$V(\Phi)$ is not a constant and we wish to deal with the usual slow-roll theory
of inflation.  Before jumping into this discussion we must say a few words about
the related question of what happens if we relax the second condition
and set $p_\phi$ to some arbitrary constant.

When $p_\phi \ne 0$ the Heisenberg equations for
the system defined by Eq.\ref{vphizero} are no longer exactly solvable; nevertheless,
we can always write the full time development operator of the theory as
\be
    U(t) = e^{i H_0 t } S(t)
\ee
where $H_0$ is the Hamiltonian of the theory obtained by setting $p_\phi=0$ and
$S(t)$ satisfies the differential equation
\be
    -i\,{d S(t) \over dt} = V_I(t) =  e^{i H_0 t } \left[ {p_\phi^2 \over 2\,u^2}\right]\,e^{i H_0 t }
                           = {p_\phi^2 \over 2\,(\cosh(\omega t) u + sinh(\omega t)p )^2} .
\ee
(i.e., $S(t)$ is given, as usual, by path ordered exponential of $V_I(t)$).
The form of $V(t)$ suggests that for an appropriate subclass of the space
of physical states (those which vanish for a region around $u_\infty=0$)
the evolution of the system is asymptotically controlled by $H_0$.  Thus, for these
states, we expect our general description of how things evolve in phase space to apply.
In any event, we see that in each sector of fixed $p_\phi$, the evolution of the system
is well defined and unitary.  Clearly, going beyond these considerations and
relaxing the condition that $V(\Phi)$ is a constant, will force us to work
with packets in $p_\phi$.  This is of course even more difficult to analyze in detail.
Thus, it is important to ask to what degree we can be sure that we will arrive
at the usual inflationary scenario, be it fast or slow-roll inflation, using this approach.
This is the question we address in the rest of this section.

Recovering the classical picture of inflation, in the limit in which our
asymptotic condition holds to high accuracy, is
straightforward.  Since we are working with the Heisenberg equations of motion,
all we have to do is assume that we start from a coherent state $\ket{\psi}$, such that
${\bf G}\ket{\psi}=0$.  Furthermore, we assume that $\bra{\psi} u \ket{\psi}$,
$\bra{\psi} p_u \ket{\psi}$, $\bra{\psi} \Phi \ket{\psi}$ and
$\bra{\psi} p_\Phi \ket{\psi}$ satisfy the initial conditions required for a
classical theory of slow-roll inflation.  In this case, it makes sense to
rewrite the Heisenberg operators as
\be
    u(t) = \widehat{u}(t) + \delta u(t) \quad;\quad
    \Phi(t) = \widehat{\Phi}(t)+ \delta\Phi(t) ,
\label{coherentapprox}
\ee
where $\hat{u}(t)$ and $\hat{\Phi}(t)$ are c-number functions
such that $\bra{\psi} u(t) \ket{\psi} = \hat{u}(t)$ and
$\bra{\psi} \Phi(t) \ket{\psi} = \hat{\Phi}(t)$.
Given these assumptions, we wish to show that if these c-number functions satisfy the
classical slow-roll equations for inflation, then as a consequence of inflation,
the quantum corrections to the Heisenberg equations of motion will be
strongly suppressed.

Substituting these definitions into the Heisenberg equation of motion for the operator $u(t)$
in Eq.\ref{heisenone} and the form of the equation of motion for the operator $\Phi(t)$ in
Eq.\ref{heisentwo},  we see that if the classical functions $\hat{u}(t)$  and $\hat{\Phi}(t)$
satisfy the classical equations for slow-roll inflation, the c-number terms all cancel,
and one is left with equations for the operators $\delta u(t)$ and $\delta\Phi(t)$.
At first glance, solving these equations seems difficult; however, the situation improves
significantly if we look at the constraint equations
\be
\left[{\cal H}^2-{\kappa^2\over 3}\left(\frac{1}{2}\left(d\Phi(t)/dt\right)^2+V(\Phi(t))
\right) + {\bf Q}(t) \right]\ket{\psi}=0,
\quad;\quad
\left[d{\cal H}/dt + {\kappa^2 \over 2} \left(d\Phi(t)/dt\right)^2\right]\ket{\psi}=0 .
\ee
In the rest of this section we will ignore the operator ${\bf Q}(t)$, since it is proportional
to  $\widehat{u}(t)^{-4}$, which we expect to be small in the inflationary and FRW eras.

The second equation says that in the sector of physical states we can replace
the operator $(d\Phi(t)/dt)^2$ by $-2(d{\cal H}/dt)/\kappa^2$.  Substituting
this in the first constraint yields the equation
\be
    \left[3{\cal H}(t)^2 + {d{\cal H}(t) \over dt} - \kappa^2 V(\Phi(t))\right]
    \ket{\psi} = 0 .
\label{errorone}
\ee
Substituting Eq.\ref{coherentapprox} into the definition of ${\cal H}$, we get
\be
    {\cal H}(t) = \widehat{{\cal H}}(t) + \delta{\cal H}(t),
\label{calH}
\ee
where the form one obtains for $\delta {\cal H}(t)$ would seem to imply that the
operator shrinks rapidly during the inflationary era because of the inverse power
of $\widehat{u}(t)$ appearing in its definition.  Nevertheless, it behooves us to
check that the operators $\delta u(t)$, {\it etc.} do not behave badly.
Substituting Eq.\ref{calH} into the constraint equation and using
$\Phi(t) = \widehat{\Phi}(t) + \delta \Phi(t)$,
cancelling out the contributions of the c-number functions and
keeping terms of first order in $\delta{\cal H}$ and $\delta\Phi(t)$,
we obtain
\be
   \left[ 6 \,\widehat{{\cal H}}\, \delta{\cal H}
    -\kappa^2 \left( {d V(\widehat{\Phi})
    \over d\Phi }\right) \delta\Phi(t) \right]\ket{\psi} = 0 .
\ee
Taking the expectation value of this equation in the state $\ket{\psi}$, noting that
by assumption $\bra{\psi} \delta\Phi(t) \ket{\psi} = 0$, it follows that
\be
    6\,\widehat{{\cal H}}(t)\,\bra{\psi} \delta{\cal H}(t) \ket{\psi} +
    {d \bra{\psi} \delta{\cal H}(t) \ket{\psi} \over dt} = 0 .
\ee
The solution to this equation is
\be
    \bra{\psi} \delta{\cal H}(t_f) \ket{\psi} = e^{-6\int_{t_i}^{t_f} dt \widehat{\cal H}(t) }
\,\bra{\psi} \delta{\cal H}(t_i) \ket{\psi} .
\label{contract}
\ee
If we recall that the classical function $\widehat{{\cal H}}(t)$ is just $d\left( \ln(a(t))\right) / dt$, we see that
the integral in Eq.\ref{contract} is just the number of {\it e}-foldings during inflation.
Thus, the contribution of the operator $\delta {\cal H}$ is strongly suppressed.

\section{How Big are the Corrections?}

Obviously, if we limit ourselves to the exactly solved case of de Sitter space, estimating
the size of the corrections to Einstein's equations in a given quantum history
reduces to calculating the ratio of the exponentially increasing term in the expression for
the volume as a function of time, to the exponentially decreasing term.
The question now is how big are the corrections to the Einstein equations in a
more general setting, and when does make sense to ignore their effects on the CMB calculation.
Another way to ask the same question is to ask whether it makes
sense to ignore the operator ${\bf Q} = -3\kappa^4/64{\bf V}^2 u^4$ at the onset of inflation.

Obviously, the issue boils down to how large this term is relative to the
operator $\kappa^2 V(\Phi)/3$.  To establish this ratio,
we must first specify the value of the quantization volume ${\bf V}$.  Clearly,
there is no upper limit for the value one can choose for ${\bf V}$.  There is, however,
a lower limit, since ${\bf V}$ must be chosen larger than the horizon volume
at the time of inflation to avoid boundary effects which are not seen in the WMAP data.
Thus, ${\bf V} > 1/{\cal H}_I^3$, where ${\cal H}_I$ is the value of the Hubble
parameter at the onset of inflation.

To estimate the size
of ${\cal H}_I$, if the classical approximation dominates,  we use the classical
version of the Friedmann equation.  This equation tells us that
\be
    {\cal H}_I^2 \approx {\kappa^2 \over 3} V(\Phi) .
\ee
Substituting this into the expression for ${\bf Q}$ we obtain that
\ba
    {\bf Q} &=& -{3\kappa^4 \over 64}\,{\cal H}_I^6 {1\over u^4} \nonumber\\
    &=& - {\kappa^{10} \over 576} V(\Phi)^3 {1\over u^4} ,
\ea
which is to be compared to $\kappa^2 V(\Phi)/3$.  Thus the statement that
${\bf Q}$ can be ignored at the onset of inflation is equivalent to
\be
    {\kappa^2 \over 3} V(\Phi) \gg {\kappa^{10} \over 576} V(\Phi)^3 {1\over u^4} .
\ee
It is convenient to multiply this equation by a factor of $\kappa^2$ to obtain
\be
    \kappa^4 V(\Phi) \gg {1 \over 192} \left(\kappa^4 V(\Phi)\right)^3 {1\over u^4} ,
\ee
or equivalently
\be
     {1\over 192} \left(\kappa^4 V(\Phi)\right)^2 {1\over u^4} \ll 1 .
\ee
At this point we note that the product $\kappa^4 V(\Phi)$ is usually
constrained to be less than or on the order of $10^{-6}$.  Thus,
if $u(t)$ is chosen to be the order of unity at
the time inflation starts, the effects of ${\bf Q}$ will be negligible.  Note however
that $1/u(t)^4 = 1/a(t)^6$, so one cannot extrapolate very many {\it e}-foldings back from
the starting point before quantum corrections become important.

\section{Remarks Concerning The Computation of CMB Anisotropy}

While we have not yet done any detailed computations, it is clear that
the fact that the quantum system deviates from pure exponential growth
at a finite time in the past could have implications for the usual
derivation of CMB fluctuations.  It is entirely possible that the
delay in the time at which the long wavelength modes of the scalar
field exit the horizon relative to the shorter wavelength modes
might produce visible effects in the predicted measurement of
$\delta \rho/\rho$.  If this is so then one should be able to
put an experimental limit on how far back in time one can
push the start of the usual computation.

\section{Possible Extensions}

In order to extend this model to a complete treatment of the CMB anisotropy,
one has both to add an extra field $\chi(t,x)$ to the metric, in order to model
Newtonian fluctuations, and to put the the spatially varying part to the $\Phi$
field back into the action.  In other words, the metric is taken to have the form
\begin{equation}
    ds^2 = -\goo\,dt^2 + a(t)^2\,\gxx\,d\vec{x}^2 ,
\end{equation}
and the action is taken to be
\be
   {\cal S} = \int d^4x \sqrt{-g}\,\left[{R(g)\over 2\kappa^2}
   + \gooinv \left(\Phidot + \dphidt \right)^2 - {\gradphisq \over \gxxinv}
   - V(\Phi(t) + \epsilon\phi(t,\x)) \right] .
\ee
If we now expand this formula up to order $\epsilon^2$
we see that: the $\epsilon^0$ term is the problem we have been considering;
the $\epsilon^1$ term vanishes due to the equations of motion; the remaining
terms are quadratic in the fields $\chi(t,x)$ and $\phi(t,x)$.
Thus, if we start in the coherent state discussed in the previous sections,
$\chi(t,x)$ and $\phi(t,x)$ are simply free fields evolving in a time-dependent background
and their Heisenberg equations of motion can be solved exactly.

This analysis allows one to completely reproduce the usual
computations for $\delta \rho/\rho$. (A complete treatment of this
will appear in a forthcoming pedagogical paper.) In other words,
this effective theory is capable of reproducing the theory of all CMB
measurements within a quantum framework in which the long
wavelength part of the gravitational field satisfies the exact Einstein equations,
while the shorter wavelengths are treated perturbatively.
The small size of the CMB fluctuations tells us this
is a reasonable approach.

This extension of the model gives us
a canonical Hamiltonian picture of the evolution of the theory. This means that
the back-reaction caused by the changes in $\chi(t,x)$ and $\phi(t,x)$ are
completely specified.  Clearly, it is important to ask if these back-reaction
effects, or the effects ${\bf Q}$ causes on the evolution of the system, leave an
observable imprint on the CMB fluctuation spectrum.

There are two less obvious but very interesting directions in which
one can extend this work.

The first is to reintroduce some very long wavelength modes of
$\chi(t,x)$ and $\phi(t,x)$ into the part of the Lagrangian
that we treat exactly. To be specific, we can expand the
field $\chi(t,x)$ and $\phi(t,x)$ in some sort of wavelet basis,
for which the low-lying wavelets represent changes over fractions
of the horizon scale; {\it i.e.\/} \be
    \chi(t,x) = \sum_{j=1}^{N} b_j(t) w_j(x) + \epsilon\chi(t,x)^{'} \quad;\quad
    \phi(t,x) = \sum_{j=1}^{N} c_j(t) w_j(x) + \epsilon\phi(t,x)^{'} .
\ee
Next, we can plug this expansion into the action and expand it to
second order in $\epsilon$.

In contrast to the previous case, the order $\epsilon^0$ part of
the action will now be the theory of $2N$ non-linearly coupled
variables.  Clearly, we should be able to parallel the discussion
given in this paper and proceed to: first, derive the canonical
Hamiltonian, then derive the Heisenberg
equations of motion for the system, which will not be the full set
of Einstein equations, and, finally, construct the proper
constraint operators to fill out the full set of equations of
motion.  The resulting theory should be equivalent to a theory in
which we have finite size boxes with independent scale factors
that are weakly coupled to one another.  Since the usual
CMB results imply that fluctuations on all wavelengths are
small, it must be true that in this version of the theory
the scale factors in neighboring boxes (or pixels) can't get very
far from one another without affecting the CMB
fluctuations.  We suggest that one way to see how
a fully quantized theory of gravity behaves at shorter wavelengths
is to pixelize the theory in this way.  Then we can study what
happens to the quantum problem as we add operators corresponding
to higher frequency fluctuations in the
quantum fields back into the fully non-linear problem.

Another interesting direction is to pixelize a problem initially
quantized in a region extending over several horizon
volumes.  Since the pixelization to volumes smaller than the scale
set by the horizon during inflation certainly leads to coupling
terms between the pixels, there will be analogous (presumably
weaker) couplings between what will now be neighboring horizon-size
pixels.  The interesting question here is whether or not the
scale factors in neighboring volumes can get very far from one
another without causing observable changes in the CMB fluctuation
spectrum seen by an observer in any individual volume.  In other
words, can we---within the context of chaotic or
eternal inflation---put experimental limits on how near to us a
very different universe from ours can be, without
leaving a visible imprint on the CMB radiation in our
universe?  If such limits can be found, we will have found a way
to see the unseeable.

\section{Summary}

In this paper we showed how to fully quantize the theory of inflation and
$\delta \rho / \rho$, at least if one takes the point of view that
getting a sensible evolution of a quantum system as a function of
cosmic time takes precedence over forcing a purely geometrical
interpretation.

Our focus throughout was on the formulation of the part of the
problem that involved the spatially constant fields.  As we
demonstrated, in both the classical and quantum theory, working in
a fixed gauge yields only two of the four relevant Einstein
equations as equations of motion.  In the classical theory we
showed that the Friedmann equation and its time derivative must be
treated as constraints whose constancy in time requires a proof.
Our proof followed from the two equations of motion we did have.
Next, we showed that, in the quantum version of the theory, the
same two Einstein equations appear as operator equations of
motion; however, surprisingly, due to quantum corrections there
were a one-parameter family of possible choices for constraint
equations.  We argued that the simplest of these constraint
equations, that which corresponds to the Wheeler-deWitt equation,
cannot be used to define the space of physical states, since it
leads to a direct conflict between the Schroedinger and Heisenberg
pictures.  However, we showed that that problem does not exist for
the other possible choices for constraint operators. Nevertheless,
we insisted that we would still run into trouble if we imposed a
strong form of any of these constraints and suggested a weaker
form of the constraint which avoids these problems.  In order to
clarify how our weaker conditions work in detail, we applied the
general formalism to the case of de Sitter (and Minkowski) space.
Our goal was to show, in an explicit, exactly solvable, case how
the formalism works.  The most important result of this discussion
is that, in the case of de Sitter space, the system deviates from
the expected pure exponential expansion at a finite time in the
past.  We also went on to discuss variants of the de Sitter
problem and then discussed the recovery of the usual inflationary
picture in the more complicated problem. One possible consequence
of the fact that the behavior of the universe at early times
differs from pure exponential expansion, is that it implies one
might measure the effects of the quantum corrections to
the pure Einstein equations as deviations from the conventionally
predicted form of $\delta \rho/\rho$. Failing that, one might bound
the earliest time at which one is free to set
initial conditions on the state of the inflaton and other fields
in the system.  To put it another way, there may either be
measurable consequences following from the quantum nature of the
problem at early times, or one will have to face up to the problem
of how and when to set initial conditions.

While, as it stands, the formalism we have presented is by no
means a candidate for a theory of everything, we feel that the
interesting results obtained by proceeding along these lines
suggests it is a good candidate for a theory of something. Namely,
a fully quantum theory of the measured fluctuations in the CMB
radiation.

\section{Acknowledgements}

We would like to thank J.~D.~Bjorken for helpful communications.

\end{document}